\newfont{\abfont}{amr10 scaled\magstep1}
\newfont{\aufont}{amr10 scaled\magstep2}
\newfont{\tifont}{ambx10 scaled\magstep3}

\def\@maketitle{\newpage   
 \null
 \vspace*{-1\headsep}      
 \vspace*{-1\headheight}
 \vspace*{-24pt}
 \begin{flushright}{\large	   
   { \preprintno} \\ \@date}
 \end{flushright}
 \vskip \headsep	   
 \vskip \headheight
 \bigskip
 \begin{center}		   
   {\LARGE\tifont \@title \par}
   \vskip 2em
   {\large\aufont
     \lineskip .5em
     \begin{tabular}[t]{c}\@author
     \end{tabular}\par}
   \vskip 1em
 \end{center}
 \par
 \vskip 1.5em}

\newcommand{\preprintno}{preprint number here}	 

\def\abstract{\if@twocolumn
\section*{Abstract}
\else \large\abfont				
\begin{center}
{\bf Abstract\vspace{-.5em}\vspace{0pt}}
\end{center}
\quotation
\fi}
\def\endabstract{\if@twocolumn\else\endquotation\fi}

\def\appendix{\par
    \setcounter{section}{0}
    \setcounter{subsection}{0}
    \renewcommand{\theequation}{\Alph{section}.\arabic{equation}}
    \setcounter{equation}{0}
}

\@addtoreset{equation}{section}
\def\theequation{\arabic{section}.\arabic{equation}}

\def\ps@columns{%
 \if@twocolumn
  \let\@mkboth\@gobbletwo
  \def\@oddhead{}\def\@evenhead{}
  \def\@oddfoot%
   {\rm\hfil\thepage\stepcounter{page}\hskip.5\textwidth\thepage\hfil}
  \let\@evenfoot\@oddfoot
 \else
 \ps@plain
 \fi
}

\documentstyle[hutp]{article}

\makeatletter \input art10.sty \makeatother
\def\starttext{\twocolumn}

\typein{*** or hit [return] for landscape mode(11x8.5) ***}

\newcommand{\beq}{\begin{equation}}
\newcommand{\eeq}{\end{equation}}

\newcommand{\remove}[1]{}
\renewcommand{\theequation}{\thesection.\arabic{equation}}

\newdimen\pmboffset
\def\oldpmb#1{\setbox0=\hbox{#1}%
 \copy0\kern-\wd0
 \kern\pmboffset\raise 1.732\pmboffset\copy0\kern-\wd0
 \kern\pmboffset\box0}

\begin{document}

\title{Double Beta Decay with Vector Majorons}

\author{
        Christopher D. Carone\thanks {carone@huhepl}  \\
        Lyman Laboratory of Physics \\
        Harvard University \\
        Cambridge, MA 02138}
\date{\today}

\renewcommand{\preprintno}{HUTP-93/A007}

\begin{titlepage}

\maketitle

\def\thepage {}        

\begin{abstract}
We consider the possibility that neutrinoless double beta decay
may occur in models with unbroken lepton number via the
emission of a massive gauge boson with electron lepton number $-2$.
We determine the shape of the $\beta\,\beta$ sum energy spectrum
as a function of the gauge coupling, independent of model-specific
details.  We discuss our results in light of the persistent experimental
claims that excess events are observed near but below the spectrum
endpoint of several elements.
\end{abstract}

\end{titlepage}

\starttext 
\pagestyle{columns} 
\pagenumbering {arabic} 

\section {Introduction} \label {sec:intro}

Double beta decay has been discussed extensively in the literature as
a potential window on physics beyond the standard model \cite{dbd}.
While two neutrino double beta decay ($2\nu\beta\beta$) occurs in the
standard model as a second order weak interaction,  neutrinoless double
beta decay ($0\nu\beta\beta$) is forbidden if one assumes that electron
lepton number is conserved.   This assumption may not be valid if the
neutrinos that participate in the charged current weak interactions have
small, lepton number violating majorana masses.  Then, the
$0\nu\beta\beta$ mode would also occur, appearing as a spike at the end
point of the electron sum energy spectrum.  If lepton number is not
conserved, then the global $U(1)$ symmetry it generates might be
spontaneously broken \cite{cmp}.  In this case,  neutrinoless double beta
decay may also proceed by the emission of the associated goldstone boson,
the majoron \cite{ggn}.  The sum energy spectrum for the majoron emitting
mode ($\beta\beta M$) is skewed toward higher energies compared to the
$2\nu\beta\beta$ spectrum,  because the available energy is shared among
fewer particles in the final state.  It has been suggested that this feature
of the $\beta\beta M$ spectrum may provide a means of explaining an
anomalous excess of events that some experimental groups have claimed to see
near but below the spectrum endpoint of several elements \cite{moe}.  In
order to account for the alleged excess without exceeding the experimental
bounds on the unobserved $0\nu\beta\beta$ mode,  one is forced in ordinary
majoron models to take the scale of lepton number breaking to be on the
order of $10$ KeV.  Since this is well below the electroweak scale
($\sim 250$ GeV), one can account for the phenomenology only at the
expense of creating an unwanted hierarchy problem.

To avoid this shortcoming, Burgess and Cline have advocated a class of
majoron models in which lepton number is exactly conserved \cite{bc}.
The `majoron' in these models is the  goldstone boson that results from
the spontaneous breaking of an extended global symmetry group $G$ that
contains electron lepton number as an unbroken subgroup.  The majoron then
carries lepton number $-2$ and the $0\nu\,\beta\beta$ mode is exactly
forbidden.  With the $0\nu\,\beta\beta$ limits automatically satisfied,
one can account for the anomalous events, while raising the symmetry
breaking scale up to $\sim 100$ MeV.  While this is still small in
comparison to the electroweak scale, it is nonetheless a $10^4$-fold
improvement over ordinary majoron models.

In this letter, we go one step further and gauge a subgroup of $G$
so that the majoron of Burgess and Cline becomes the longitudinal
component of a massive gauge boson.  Double beta decay can proceed
via the emission of these new gauge bosons, which carry lepton number $-
2$.   If we take the gauge coupling to zero, we should recover the results
of ref. \cite{bc} (with the exception that our model will have a
noninteracting massless vector meson remaining.)  We will determine how
the shape of the $\beta\,\beta$ spectrum is effected as one turns on
the gauge coupling, and discuss how this effects the relevant
phenomenology.

\section {The Sum Energy Spectrum} \label {sec:sumspec}

In this section, we determine the shape of the sum energy spectrum for
double beta decay with vector majoron emission ($\beta\beta X^\mu$).
To avoid conflict with the measured width of the $Z$, we assume, in
the spirit of ref. \cite{bc}, that the new gauge field $X^\mu$
is an electroweak singlet.  It participates in double beta decay by
coupling to electroweak singlet neutrinos that mix with the gauge
eigenfield $\nu_{e\,L}$.  To provide an example, let us consider
the `charged majoron' model suggested in ref. \cite{bc}:  The standard
model is augmented by the global symmetry group
$G=SU(2)_s \times U(1)_{L'}$, where electron lepton number is the $U(1)$
subgroup generated by $L_e = 2 \, T_3 + L'$.  The model includes nonstandard
electroweak singlet majorana neutrinos $N=(N_+\mbox{,}N_-)$, $s_+$,
and $s_-$, whose left-handed components transform as (2,0), (1,1), and (1,-1)
under $G$, respectively.  A singlet majoron, transforming as a (2,1),
couples to these neutrinos and to the usual electron neutrino,
$\nu_{e\,L}$, which transforms as a (1,1).  When $G$ spontaneously breaks,
the resulting neutrino mass eigenstates participate in the
weak interactions through their $\nu_e$ component, and couple to the
majoron, through their $N$ component.  Thus, a suitable example of the
vector majoron models of interest to us here is obtained by gauging
the $SU(2)_s$ in the model just described.  The majoron becomes the
longitudinal component of the now massive $SU(2)_s$ gauge boson, $X^\mu$,
which takes part in the neutrinoless $\beta\beta X^\mu$ decay mode.

To describe the spectrum of neutrinos in models of this type, we work in
a basis of majorana mass eigenstates, $\nu_i$.  The electron neutrino
$\nu_e \equiv \nu_{e\,L} + \nu_{e\,L}^c$ can then be expressed in
terms of the mass eigenstates
\beq
\nu _e \, = \, \sum_{i} V_{e\,i} \, \nu _i
\label{eq:nudef}
\eeq
where $V_{ab}$ is the neutrino mixing matrix.  Next, we assume that the
vector majoron-neutrino coupling is of the form
\beq
{\cal L} \,=\, -\frac{1}{2} \overline{\nu_i} \gamma_\mu
(c_{ij} P_L +d_{ij} P_R) \nu_j \, X^\mu
\label{eq:coupdef}
\eeq
where $c_{ij}$ and $d_{ij}$ depend on the details of the model.  Using the
fact that the $\nu_i$ are majorana fields ($\nu_i = \nu_i^c$) it is
straightforward to show that the couplings $c$ and $d$ are related,
\beq
c_{ij} \, = \, - d_{ji}
\label{eq:relate}
\eeq
Equation (\ref{eq:relate}) assures that there are no vector
couplings between majorana fields of the same flavor; this must be the
case, or else one would be able to construct an operator out of
these fields that counts a conserved lepton number.

Equations (\ref{eq:nudef}) and (\ref{eq:relate}) allow us to write down
the leptonic part of the double beta decay Feynman diagrams.  To evaluate
the hadronic part, we adopt the approximation scheme of ref. \cite{bc};
we neglect the 4-momenta of the electrons and the vector majoron, $p_1$,
$p_2$ and $q$ respectively, and parameterize the nuclear matrix element
solely in terms of the loop momentum $p$, and the four-velocity of the
nuclei $u$.  Thus,
\begin{eqnarray}
\lefteqn{ \frac{1}{\sqrt{2m_n}}\frac{1}{\sqrt{2m_{n'}}}
\int d^4 x d^4y < N'| J^\rho (x) J^\sigma (y) | N > e^{ip\cdot x} e^{-i
p\cdot y} = } \nonumber \\
& & w_1 g^{\rho\sigma} + w_2 u^\rho u^\sigma + w_3 (p^\rho
u^\sigma-p^\sigma u^\rho) + w_4 \epsilon ^{\rho\sigma\alpha\beta}
u_\alpha p_\beta + w_5 p^\rho p^\sigma
\label{eq:nucpart}
\end{eqnarray}
\[
\equiv W^{\rho\sigma}
\]
where $m_{n}$ and $m_{n'}$ are the masses of the initial and final
nuclei.  Notice that we are working in the recoilless approximation; the
4-velocity of the initial nucleus remains effectively unchanged by the
small momentum transfers involved in the decay.  Neglecting the dependence
of (\ref{eq:nucpart}) on the 4-momenta of the outgoing particles is
a reasonable approximation because the loop momentum $p$ is cut off at
the nuclear Fermi momentum, $p_F \approx 100$ MeV, while
$p_1$, $p_2$, and $q$ are on the order of the end point energy,
$Q \approx 1$ MeV.  Finally, we only retain terms that are
lowest order in the vector majoron 4-momentum.  For example, the
denominators of the neutrino propagators can be expanded
\beq
\frac{1}{(p\pm q/2)^2-m_{\nu_i}^2} = \frac{1}{p^2-m_{\nu_i}^2}
\left[1\mp \frac{p\cdot q}{p^2 - m_{\nu_i}^2} + \cdots\right]
\label{eq:expand}
\eeq
The second term in brackets gives a contribution to the amplitude
that is smaller than the first by a factor of $Q/p_F \approx 1$\%.

In unitary gauge, the spin-summed squarred amplitude will have the
form
\beq
<|{\cal A}|^2> = M^{\mu\nu} \sum_{rs} \epsilon^{(r)}_\mu
\epsilon^{(s)\,*}_\nu = M^{\mu\nu} \left( - g_{\mu\nu} + \frac{q_\mu q_\nu}
{M^2}\right)
\label{eq:theform}
\eeq
where $\epsilon_\mu^{\,(a)}$ are the polarization vectors for the $X^\mu$
boson, and $M$ is its mass.  Notice that $M^{\mu\nu}$ is proportional
to $g^2$, where $g$ is the gauge coupling.  Hence, the term proportional
to $-g^{\mu\nu}$ vanishes in the $g\rightarrow 0$ limit, but the second
term remains finite since $M^2 \propto g^2$. In the $g\rightarrow 0$ limit,
the surviving term should reproduce the charged majoron amplitudes presented
in ref. \cite{bc}.

Before writing down the answer, it is instructive to provide one
intermediate step.  If we seperate the nuclear matrix element
$W^{\rho\sigma}$ into pieces that are symmetric or antisymmetric under
the interchange $\rho\rightleftharpoons\sigma$
then after some algebra, we can write the amplitude as
\[
{\cal A} = \sum_{ij} V_{ei} V_{ej} \int \frac{d^4 p}{(2\pi)^4}
\frac{1}{(p^2 - m_{\nu_i}^2)} \frac{1}{(p^2 - m_{\nu_j}^2)}
\, 4\epsilon_\mu^{\,*}
\]
\[
\times \left[ 2g_{\rho\sigma} W^{\rho\sigma}_{(s)} p^\mu \overline{u}_1
P_R u_2^c (c_{ij}m_j+m_i d_{ij}) \right.
\]
\beq
\left.
+W^{\rho\sigma}_{(a)} \overline{u}_1 \gamma_\rho [\not\!p\mbox{,}\gamma^\mu]
\gamma_\sigma P_R u_2^c (c_{ij}m_j-m_i d_{ij}) \right]
\label{eq:intstep}
\eeq
The first term in brackets vanishes identically when we sum over
$i$ and $j$ because $c_{ij}m_j+m_i d_{ij}$ is antisymmetric under
$i \rightleftharpoons j$, while what multiplies it is symmetric
(c.f. eqn.(\ref{eq:relate})).  The result, therefore, depends only
on the nuclear form factors $w_3$ and $w_4$.

We are now in a position to state the result.  After integrating over
the appropriate region of phase space, the partial decay with $d\Gamma$
can be written
\beq
d\Gamma = \left(\frac{G_F}{\sqrt{2}}\right)^4 |\,{\cal A}\,|^2 \, d\Omega
\eeq
where
\beq
d\Omega = \frac{1}{64\pi^5}
\left[(Q-\epsilon_1-\epsilon_2)^2-M^2\right]^{\frac{3}{2}}
\prod_{k=1}^{2} |\vec{p}_k| \epsilon_k F(\epsilon_k, Z) d\epsilon_k
\label{eq:dom}
\eeq
and where ${\cal A}$ is given by
\beq
{\cal A} =
8\sqrt{2} \sum_{ij} V_{ei} V_{ej} b_{ij}
\int \frac{d^4 p}{(2\pi)^4} \frac{2 \vec{p}\,^2 (w_3 - i w_4)}
{(p^2-m_{\nu_i}^2)(p^2-m_{\nu_j}^2)} + {\cal O}(g^2) +\cdots
\label{eq:rest}
\eeq
Above, the Fermi functions $F(\epsilon_k, Z)$ incorporate the Coulomb
corrections due to the charge of the final nucleus, with atomic number $Z$.
Notice that we have arbitrarily collected all the dependence on the
electron energies into the quantity $d\Omega$, which also includes the
4-body phase space factor. We have defined the symmetric matrix $b_{ij}$
\beq
b_{ij} \equiv \frac{(c_{ij} m_j - m_i d_{ij})}{2 M}
\eeq
to make the connection between our result, and that of ref. \cite{bc}
clear; in the $M\rightarrow 0$ limit, we should recover the correct
charged majoron amplitude\footnote{The astute reader might notice that
our result does not agree with the amplitude published by Burgess and
Cline, but this is due to some errors in the early portion of
their paper;  they inadvertantly omit one of the two Feynman
diagrams, the one with the electron lines crossed,  and thereby generate
a spurious term proportional to $W^{\rho\sigma}_{(s)}$.
In addition, the portion of their result containing the correct
form factors displays them in the combination $w_3-2 i w_4$,
which also is the result of an algebraic error.  We thank C. Burgess
for confirming these points.  Finally, the normalization of our result
disagrees with theirs.  According to one of the authors, these errors do
not effect the subsequent numerical estimates presented in ref. \cite{bc},
which were based on amplitudes computed independently by the
other author.}.  The conclusion of our analysis is that
the $(Q-\epsilon_1-\epsilon_2)^3$ dependence of the charged majoron result
becomes $[(Q-\epsilon_1-\epsilon_2)^2 - M^2]^{3/2}$
in the case where the majoron is eaten.  This can be understood as a
consequence of the vanishing of eqn. (\ref{eq:intstep}) when the
index $\mu\neq 0$, as one can confirm by manipulating the expression using
the explicit form of $W^{\rho\sigma}_{(a)}$.  As a result, the sum over
vector spin states yields an overall factor of $(-g^{00}+(q^0)^2/M^2)$,
which accounts for the energy dependence of (\ref{eq:dom}).  The variation
in shape of the $\beta \beta$ sum energy spectrum as a function of the
vector majoron mass is displayed in Figure 1.

\section {Discussion} \label {sec:discuss}

The most important (and obvious) point about the sum energy spectrum
for the $\beta\beta X^\mu$ mode is that it has a smaller endpoint energy
than the other $\beta\beta$ modes, as a consequence of
the mass of the vector majoron.  Shifting the endpoint of the anamolous
spectrum to smaller energies is not the most desirable feature of gauging
the model, if we are attempting to explain excess events
on the tail of the $\beta\beta$ spectrum.  Nevertheless, this class
of models remains viable providing that the vector majoron mass
is sufficiently small.  For example, if we take the symmetry breaking
scale to be $v\sim 100$ MeV, for the reasons discussed in ref. \cite{bc},
and the gauge coupling $g$ to be on the order of $10^{-4}$, then the
vector majoron mass is of order $10$ KeV, and the endpoint of the anomalous
spectrum is shifted by this amount.  This shift is small, however,
compared to the $\sim 1$ MeV interval in which the extra counts
are observed.  With such a small gauge coupling the $\beta\beta X^\mu$
decay rate does not differ appreciably from the charged majoron rates
discussed in ref. \cite{bc}, which are of the appropriate size to
account for the observed excess.  For $g\sim 10^{-4}$, the vector majoron
rate is only $10$ \% less than the charged majoron rate at a sum electron
energy $\sim 40$ KeV below the endpoint.  If the anomalous events are
ever confirmed experimentally, and the if their spectrum has the features
predicted in this letter, than a more detailed analysis, involving
estimates of the nuclear form factors, would be required to determine
whether the vector majoron mode yields the correct anomalous rate.

\vspace{36pt}
\centerline{\bf Acknowledgments}
I thank Howard Georgi, as always, for being an endless reservour
of interesting research ideas, and Thorsten Ohl for
introducing me to FORM.  I also thank C. Burgess and J. Cline for
clarifying their results. {\it This work was supported in part by
the National Science Foundation, under grant PHY-87-14654, and
the Texas National Research Laboratory Commission, under grant
RGFY9106.}


\begin{center}
{\bf Figure Caption}
\end{center}
{\bf Fig. 1.}  The electron sum energy spectrum for the vector majoron
emitting mode (ignoring Coulomb corrections).  The solid, dashed, and
dot-dashed spectra correspond to vector majoron masses of $M = 0$, $250$, and
$500$ KeV, respectively.  The normalization factor, $\Gamma_0$, is the
full width of the $M=0$ decay mode.

\end{document}